\documentclass{nature}
\usepackage{epsfig}
\usepackage[10pt]{moresize}
\usepackage{comment}
\usepackage{amsmath}
\usepackage{amssymb}
\usepackage{accents}

\begin{document}

\title{Plasmonic bio-sensing for the Fenna-Matthews-Olson complex}
\author{Guang-Yin Chen$^{1,\dag}$, Neill Lambert$^{2}$, Yen-An Shih$^{3}$, Meng-Han Liu$^{3}$, Yueh-Nan Chen$^{3,4\star}$, and Franco Nori$^{2,5}$}
\maketitle

\begin{affiliations}
\item
Department of Physics, National Chung Hsing University, Taichung 402, Taiwan
\item
Center for Emergent Matter Science, RIKEN, Wako-shi, Saitama 351-0198, Japan
\item
Department of Physics, National Cheng-Kung University, Tainan 701, Taiwan
\item
Physics Division, National Center for Theoretical Sciences, Hsinchu, Taiwan
\item
Physics Department, University of Michigan, Ann Arbor, Michigan 48109-1040, USA
\\$^\star$e-mail:gychen@phys.nchu.edu.tw
\\$^\star$e-mail:yuehnan@mail.ncku.edu.tw
\end{affiliations}

\begin{abstract}
We study theoretically the bio-sensing capabilities of metal nanowire surface plasmons.  As a specific example, we couple the nanowire to specific sites (bacteriochlorophyll) of the Fenna-Matthews-Olson (FMO) photosynthetic pigment protein complex. In this hybrid system, we find that when certain sites of the FMO complex are subject to either the suppression of inter-site transitions or are entirely disconnected from the complex, the resulting variations in the excitation transfer rates through the complex can be monitored through the corresponding changes in the scattering spectra of the incident nanowire surface plasmons. We also find that these changes can be further enhanced by changing the ratio of plasmon-site couplings. The change of the Fano lineshape in the scattering spectra further reveals that ``site 5" in the FMO complex plays a distinct role from other sites. Our results provide a feasible way, using single photons, to detect mutation-induced, or bleaching-induced, local defects or modifications of the FMO complex, and allows access to both the local and global properties of the excitation transfer in such systems.
\end{abstract}

Photosynthesis, the transformation of light into chemical energy, is one of the most crucial bio-chemical processes for life on earth \cite{van}. When a light-harvesting antenna absorbs photons, the resulting electronic excitation is transferred to a reaction center where it is transformed into other types of energy. With its relatively small size, homogenous structure, and solubility, the Fenna-Matthew-Olson (FMO) complex in green sulfur bacteria has attracted much research attention and has been widely studied \cite{Blankenship} as a prototypical example of a photosynthetic complex. It consists of eight sites (chromophores), each of which can be regarded as an effective two-level system with flourescent resonant energy transport coupling between one another \cite{aki, Lambert, flo, Sch, Gosh1, Gosh2}. The FMO complex is surrounded by a protein environment, which normally leads to decoherence and noise, but is widely thought to play a role in assisting the excitation transfer in the complex \cite{aki, model1, model2, model3, NL2, NL3}. The excitation transfer in the FMO complex was first demonstrated \cite{Fleming} to exhibit signatures of non-negligible quantum coherence at 77 K, and then more recently advanced to room temperature \cite{RT1, RT2}.

Recently, hybrid quantum systems, which combine two or more physical systems, enable one to combine the strengths and advantages of individual systems. Work in this area has led to new phenomena and potentially new quantum technologies \cite{hybrid}. Inspired by this approach, and motivated by recent developments on achieving interactions between surface plasmons (SPs) and organic molecules \cite{nanoletter, ACSnano, Ja}, here we investigate a hybrid system which integrates nanowire SPs with the FMO complex, with the goal of probing properties of the complex via the surface plasmon scattering spectra.  Such SPs are electromagnetic excitations existing on the surface of metals \cite{rev3}, which can be excited by external fields. Due to their strong interactions with emitters, significant enhancement of the atomic or excitonic decay rates have been observed \cite{rev3, Akimov, gy, gyPRA, Chang, Chang2, en_nm, en_APL}. With strong analogies to light propagation in conventional dielectric components \cite{Zia, savel}, nanowire SP have been used to achieve subwavelength waveguiding below the diffraction limit, bipartite quantum entanglement \cite{PRB_GY, enta}, and the miniaturization of existing photonic circuits \cite{Boz}. The strong coupling between the emitters and SP fields also enables the system to act like a lossy optical cavity; namely, the interaction can be coherent \cite{Chang2, gyPRA}. These advantages of strong coupling between SPs and emitters and the relatively low-power propagation loss of SP in the nanowire make nanowire SPs an attractive system to combine with the environment-assisted transport in the FMO complex, for the goal of enhanced bio-sensing. While challenging, the recent breakthroughs on coupling SPs to J-aggregates \cite{ACSnano, Ja} and Photosystem I trimer complexes \cite{nanoletter}  suggests the hybrid devices we study here may be feasible in the future.

Like all other species, the \textit{in vivo} FMO complex can experience both mutation-induced and bleaching-induced local defects or non-functional sites, leading to blocked excitation transferring pathways. This has been demonstrated in recent experiments \cite{mutation1, mutation2}. These effects may be observed in the changes in excitation dynamics, the efficiency of excitation arrival at the reaction center \cite{JMCP}, or the spectra of the photoluminescence \cite{nanoletter, ACSnano}.  Still lacking however is both a means to prepare the complex with a single localized excitation, and a means to measure populations of local sites of the complex. Here we examine how the SP can be used as both a single photon source \cite{Akimov} and a detector. Through the scattering of the incident SP due to the couplings between the SPs and sites 1 and 6 of the FMO complex, the changes in the transmission spectra  can indicate the presence of pathway-inhibition or missing sites. This provides an alternative method to detect mutation-induced defects or non-functional sites, and suggests further applications for such hybrid systems.
\section*{Results}	
We model a single FMO monomer as a network of $N=7$ sites (see Fig.~1), which can be described by a general
Hamiltonian
\begin{equation}
H_{\textrm{FMO}}=\sum_{n=1}^{N}\epsilon _{n}|n\rangle \langle n|+\sum_{n<n^{\prime
}}J_{n,n^{\prime }}(|n\rangle \langle n^{\prime }|+|n^{\prime }\rangle
\langle n|)
\end{equation}%
where the state $|n\rangle $ represents an excitation at site $n$ ($n\in $
1,...,7), $\epsilon _{n}$ is the site energy of chromophore $n$, and $%
J_{n,n^{\prime }}$ is the excitonic coupling between the $n$-th and $n^{\prime }$%
-th sites. For simplicity, here the recently discovered \cite{olbrich} eighth site has been omitted because
results of molecular dynamics simulations \cite{olbrich} suggest that this site plays a minimal role in the processes we are interested in.

In the bacterial photosynthesis, the excitation from the light-harvesting antenna enters the FMO complex at sites 1 or 6 and then transfers from one site to another. When the excitation gets to the site 3, it hops irreversibly to the reaction center.  In the regime
that the excitonic coupling $J_{n,n^{\prime }}$ is large compared with the reorganization energy, the electron-nuclear coupling
can be treated perturbatively \cite{aki}, and the dynamics of the system can be governed by the quantum Liouville equation. The strong coupling between the excitonic dynamics and the FMO environment, as well as the structure of that environment, will quantitatively affect the excitation transfer ~\cite{ishizaki, muroukh}. Here, however, we wish to focus on the interplay between the excitation transfer and the SP scattering; so we focus on the simplest possible model of such environmental effects.  This allows us to clearly identify in what way defects in the FMO complex affect the SP scattering. A full investigation of the environment influence will be considered in future work. Also, it should be noted that excitonic fluorescence relaxation is not included in the Liouville equation. This is because its time scale ($\sim 1$ ns) is much longer compared with that of the excitation transfer from site 3 to the reaction center ($\sim 1$ ps), the typical excitation transfer time across the complex, and the dephasing \cite{seth} ($\sim 100$ fs), such that this relaxation process can then be omitted for simplicity.

%
%\begin{equation}
%\dot{\rho}(t)=-\frac{i}{\hbar }[H,\rho ]+\mathfrak{L}[\rho ],
%\end{equation}%
%where $\rho $ is the system density matrix, and $\mathfrak{L}[\rho ]$ denotes the
%Lindblad operators
%\begin{equation}
%\mathfrak{L}[\rho ]=\mathfrak{L}_{\text{sink}}[\rho ]+\mathfrak{L}_{\text{deph}}[\rho ].
%\end{equation}%
%Here, $\mathfrak{L}_{\text{sink}}$ describes the irreversible excitation transfer from
%site 3 to the reaction center:
%\begin{equation}
%\mathfrak{L}_{\text{sink}}[\rho ]=\gamma_{\textrm{s}} \lbrack 2\hat{s}\rho \hat{s}^{\dagger }-\hat{s}^{\dagger }\hat{s}\rho
%-\rho \hat{s}^{\dagger }\hat{s}],
%\end{equation}%
%where $\hat{s}=|0\rangle \langle 3|$, with $|0\rangle $  plays the role of the state of the
%reaction center, and $\gamma_\textrm{s} $ the transfer rate. The other Lindblad
%operator, $\mathfrak{L}_{\text{deph}}$, describes the temperature-dependent dephasing
%with rate $\gamma _{\text{dp}}$:
%\begin{equation}
%\mathfrak{L}_{\text{deph}}[\rho ]=\gamma _{\text{dp}}\sum_{n}[2\hat{A}_{n}\rho \hat{A}_{n}^{\dagger
%}-\hat{A}_{n}\hat{A}_{n}^{\dagger }\rho -\rho \hat{A}_{n}\hat{A}_{n}^{\dagger }],
%\end{equation}%
%where $\hat{A}_{n}=|n\rangle \langle n|$, is a projection operator into site $n$. This dephasing Lindblad operator results
%in exponential decay of the quantum coherence in the system density matrix. Typically it is thought that this perturbative treatment of the environment breaks down, and more sophisticated techniques should be employed~\cite{ishizaki, muroukh}.

\textbf{The Surface plasmon and FMO Hybrid system.}
As shown in Fig.~1, we consider a metal nanowire that is placed close to the FMO complex, as a substitute for the light-harvesting antenna found {\em in vivo}. A SP with energy $E_k=\hbar v_g k$ incident from the left end of the wire can be strongly coupled \cite{APL, ncomm, nanoletter} to both sites 1 and 6.  Here, $v_g$ and $k$ are the group velocity and wave vector of the incident SP, respectively, and $v_g$ is set to be unity throughout this paper. The incident SP can then be scattered by the two sites, due to these strong SP-site couplings. Alternatively, it can be absorbed by these two sites and be dissipated due to loss in the FMO complex. The total Hamiltonian $H_\textrm{T}$ of this SP-FMO hybrid system can be written as \cite{PRB_GY, Shen2, zhou} $H_{\textrm{T}}=H_{\textrm{sp}}+H_{\textrm{sp-site}}+H_{\textrm{FMO}}$, with
\begin{eqnarray}
&H_{\textrm{sp}}&=\int \!\!dk\; \hbar\omega_k\; a^{\dagger}_ka_k\notag\\
&H_{\textrm{sp-site}}&=\int \!\!dk \;\hbar \left[(g_{1}\sigma^{+}_1a_k+g_{6}\sigma^{+}_6 a_ke^{ikd})+\textrm{H.c.}\right],\notag\\
\end{eqnarray}
where $H_{\textrm{sp}}$ stands for the energy of the SPs with $a^{\dag}_k$ being the creation operator of the $k$-mode SP, $H_{\textrm{sp-site}}$ denotes the interaction between the SPs and the sites of the FMO complex. In $H_{\textrm{sp-site}}$, $d$ is the separation between site 1 and 6, while $\sigma^+_{1(6)}$ denotes the raising operator for site 1 (6). Here, the coupling strength $g_{1(6)}$ between the SPs and site 1 (6) is assumed to be independent of $k$ under the Markovian approximation \cite{Shen2,Gardiner}. The strong decay rate into SPs  then takes the form \cite{Chang2}, $\gamma_{\textrm{sp}}=4\pi g_{1(6)}^2/(d\omega_k/dk)$, and the values we choose for $g_1$ and $g_6$ are consistent with a recent analysis of nanowire SPs coupled to J-aggregates \cite{nanoletter, ACSnano, Ja}.

When the SP is propagating on the surface of the metal nanowire, it inevitably suffers from dissipation, such as Ohmic loss, that can be described by a Markovian channel. Thus, here we include the Markovian channels for irreversible excitation transfer from site 3 to the reaction center, as well as the Ohmic loss, by introducing non-Hermitian terms in the total Hamiltonian. Moreover, since we are interested in time-independent solutions of the photon state, the effect of quantum jumps \cite{jump} on the dynamics is neglected, such that the phonon dephasing leads to population reduction in our model. We therefore simply include the phonon dephasing as a non-Hermitian term. While this introduces extra population loss, it allows us to include in a simple way the broadening effect of dephasing on the SP spectra. The site energy of the FMO complex in Eq.~(1) is then modified as
\begin{equation}
\sum_{n=1}^{N=7}\Bigl(\epsilon _{n}-i\frac{\gamma_n}{2}\Bigr)|n\rangle \langle n|,
\end{equation}
where $\gamma_n$ are mostly zero, except for the rates referring to the Markovian channels, $\gamma_{1}=\gamma_{6}=\gamma_{\textrm{dp}}+\gamma_{\textrm{ol}}$ and $\gamma_3=\gamma_{\textrm{s}}$. Here, the Ohmic loss rate $\gamma_{\textrm{ol}}$ is set to be $\gamma_{\textrm{ol}}=20^{-1}\gamma_{\textrm{sp}}$ \cite{rev3, Chang2, NM}. Note that the separation between site 1 and 6 is about 1-2 nm \cite{size}, which is much smaller than the wavelength of the incident SP. Therefore, one can set the separation $d=0$. Similarly, since the separation is small compared with the distance ($\sim$ 30 nm) that SPs propagate during the dephasing time of scattering \cite{mfp, edp}, we neglect plasmon dephasing. Note that we only include $\gamma_{dp}$ on sites $1$ and $6$, and neglect dephasing for other sites, because the broadening of these sites, which couple to the SPs, typically dominates the scattering spectra \cite{PRL_dp}.

The energy eigenstate of the hybrid system with an energy matching the incident SP, $E_k = v_gk$, can be written as \cite{Chang2, Shen2}:
\begin{eqnarray}
\label{Ek}
|E_k\rangle&=&\int dx\left[\phi(x)^{\dagger}_{k,R}C^{\dagger}_{R}(x)+\phi^{\dagger}_{k,L}(x)C^{\dagger}_{L}(x)\right]|\textrm{g},0_{\textrm{sp}}\rangle\notag\\
	&+&\sum^7_{n=1}\xi_n|n,0_{\textrm{sp}}\rangle
\end{eqnarray}
where $|\textrm{g},0_{\textrm{sp}}\rangle$ describes that the FMO complex is in the
ground state with no SPs, and $|n,0_{\textrm{sp}}\rangle$ stands for that the excitation is in the site $n$, while $\xi_n$ is the probability
amplitude that the site $n$ absorbs the excitation. We also assume that the SP field is incident from the left of the waveguide,
the scattering amplitudes $\phi_{k,R}^{\dag }(x)$ and $\phi_{k,L}^{\dag }(x)$ therefore take the form,
\begin{eqnarray}
\phi_{k,R}^{\dag }(x)&\equiv& [\textrm{exp}(ikx)\theta (-x)+t~\textrm{exp}(ikx)\theta (x)],\notag\\
\phi _{k,L}^{\dag }(z)&\equiv& r~\textrm{exp}(-ikx)\theta(-x).
\end{eqnarray}
Here, $t$ and $r$ are the transmission and reflection amplitudes, respectively, and $\theta (x)$ is the unit step function. The
total Hamiltonian $H_\textrm{T}$ can be further transformed \cite{Shen2} by Fourier transformation into a real-space representation,
\begin{eqnarray}
&&\tilde{H}_\textrm{T}=\hbar \int dx\left\{-iv_{\textrm{g}}c_{R}^{\dag }(x)\frac{\partial }{\partial x}%
c_{R}(x)+iv_{\textrm{g}}c_{L}^{\dag }(x)\frac{\partial }{\partial
x}c_{L}(x)\right. \notag
\\
&&\left. +\hbar g_1\delta (x)\left[c_{R}(x)\sigma ^+_{1}+c_{L}(x)\sigma^+_{1}+\textrm{H.c.}\right]\right.\notag\\
&&\left. +\hbar g_6\delta (x-d)\left[c_{R}(x)\sigma ^+_{6}+c_{L}(x)\sigma^+_{6}+\textrm{H.c.}\right]\right\} \notag \\
&&+\sum_{n=1}^{N}\Bigl(\epsilon _{n}-i\frac{\gamma_n}{2}\Bigr)|n\rangle \langle n|+\sum_{n<n^{\prime
}}J_{n,n^{\prime }}(|n\rangle \langle n^{\prime }|+\textrm{H.c.})
\end{eqnarray}
where $c_{R}(x)$
[$c_{L}(x)$] is a bosonic operator annihilating a right-going
(left-going) photon at $x $, and $\delta(x)$ is the Dirac delta function. This real-space total Hamiltonian can be applied to the energy eigenstate [Eq.~(\ref
{Ek})]. The transmission spectrum $T=\left| t\right| ^{2}$ and the probability amplitudes $\xi_n$ can then be obtained by solving the eigenvalue equation $\tilde{H}_T|E_{k}\rangle =E_{k}|E_{k}\rangle $. Note that in the following calculations, we employ the energies and excitonic couplings from Ref.~[50] for the FMO Hamiltonian. The rate of excitation transfer from site 3 to the reaction center is set to be $\gamma^{-1}_{\textrm{s}}=1$ ps, and the dephasing rate $\gamma_{\textrm{dp}}$, proportional to the temperature \cite{seth}, is chosen to be 77 cm$^{-1}$.

\textbf{Detecting defects from changes in the scattering spectra.}
It has been demonstrated experimentally \cite{mutation1, mutation2} that each chromophore in the FMO complex can be decoupled from its nearest site due to rotations of the chromophore or mutation-induced local defects. This can lead to damage of the chromophores, such as blocked energy pathways or entirely non-functional sites. Here we first calculate the transmission spectra of the incident SP field in the presence of defects in the FMO complex, and compare them to the case without any defects.  As can be seen in Fig.~2, the black-solid curve shows the transmission spectrum of the incident SP field for the normal FMO complex. The dips
in the spectrum correspond to the the eigenenergies of the hybrid system, i.e., the dips occur when the sites resonant with the SP reflect the incident field \cite{Chang2}. The reason why the dips do not reach zero is because
here we have included  dephasing and the irreversible excitation transfer from site 3 to the reaction center. These dissipative channels  decrease the amplitude of the dips. As seen in Fig.~2,
the red-dashed curves show the transmission spectra when certain excitonic couplings are inhibited. The differences between the red-dashed and black-solid curves can indicate the occurrence of a blocked transfer pathway. However, the spectral differences in Fig.~2(a) and (d) are larger than those in Fig.~2(b) and (c). This is because in the normal case these sites are strongly coupled, and thus when suppressed  the inhibition strongly affects not only the eigenenergies, but also the quantum coherence \cite{CM} in the site basis. On the other hand, the transmission spectra in Fig.~2 show typical Fano lineshapes \cite{Fano} stemming from the interference between discrete (the sites) and continuous channels (the SPs).  As the inhibition of the relatively strong coupling $J_{5,6}$, between sites $5$ and $6$ leads to the enhancement of the coherence between other the sites of the FMO complex, the Fano resonance is strongly altered [Fig.~2(d)]. This result is consistent with our previous work \cite{PRE} on the dynamics of the excitation transfer in FMO complex.

Similarly, mutation-induced defects or local environment modification \cite{mutation1, mutation2} can also lead to the disconnection of the chromophores from the FMO complex entirely, rather than just suppression of certain couplings. In Fig.~3, we plot the transmission spectra for the cases where certain sites with relatively large excitonic couplings are removed entirely. The differences between the black-solid (the normal situation) and red-dashed curves can again indicate this drastic alteration of the complex. One interesting question can be raised here, can one distinguish the following two situations: (i) inhibiting a strong excitonic coupling, and (ii) totally removing one of the sites which contains the very excitonic coupling entirely? The answer is yes. By comparing Fig.~2 and 3, the inhibition of the transfer pathway only results in changes of the transmission spectra, but the missing site case also leads to vanishing peaks/dips, indicating the removal of an eigenstate of the system. Note that in plotting Fig.~2 and 3, we assume the SP-site couplings to be symmetric, i.e., $g_1=g_6=10$ cm$^{-1}$. In the following section, we will discuss the effects of asymmetric couplings.
\section*{Discussion}
So far we have shown that local defects or missing sites can be detected via changes in the transmission spectra of the incident SP.
However, if these changes in the transmission spectra can be enhanced, then the observations of such an effect in experimental realizations would become more feasible. We have observed that the transmission spectra contain
 Fano lineshapes, thus we can predict that a  discrepancy between the discrete (the FMO sites) and continuous (the SPs) channels will strongly affect the behavior of such Fano resonances \cite{PRA, OE, Fano2}. Larger changes in the transmission spectra  can then be achieved by making these two channels more disparate. In Fig.~4, we show the transmission spectra when inhibiting $J_{1,2}$
[Fig.~4(a)] and entirely removing the site 2 [Fig.~4(b)] with asymmetric SP-site couplings $g_1/g_6=100$ ($g_1=10$ cm$^{-1}$, $g_6=0.1$ cm$^{-1}$). The changes in the transmission spectra are enhanced compared to
those in Fig.~2(a) and Fig.~3(a), respectively. This is because when the damage (pathway-inhibition or missing site) occurs to site 2, the coherence between other sites in the FMO complex increases \cite{PRE},  and  the large ratio $g_1/g_6$ reduces the communication between discrete and continuous channels. As a result, the two channels become more disparate, leading to an enhancement in the transmission spectra.

Similar results can also be obtained as shown in Fig.~5 by either inhibiting $J_{5,6}$ [Fig.~5(a)] or completely removing site 5 [Fig.~5(b)]. Correspondingly, the asymmetric couplings here are chosen as inverse of the previous example $g_1/g_6=0.01$ ($g_1=0.1$ cm$^{-1}$, $g_6=10$ cm$^{-1}$). Again we see that the changes in the transmission spectra are enhanced compared with those in Fig.~2(b) and Fig.~3(b). More importantly, in Fig.~5, we can see that when
site 5 is disconnected from the FMO complex, the sharp Fano lineshape on the right-hand side of Fig.~5 is smeared out. Since $J_{1,2}$ ($-$104.1 cm$^{-1}$) is larger than $J_{5,6}$
(89.7 cm$^{-1}$), we would expect this effect to be obvious in Fig.~4, but it is not. This reveals that site 5 is very distinct from other sites, consistent with our previous work  \cite{PRE}.

Since the FMO protein complex is in reality a trimer, the validity of the results obtained by studying the monomer alone shall be addressed here. Typically studies have shown that
the subunits of the trimer act as separate transport channels \cite{olbrich} from the antenna to the reaction center. On the other hand, the site-couplings of a monomer were found to be increased because of the other monomers in the trimer system \cite{Ritschel}. This suggests that studying individual FMO monomers is sufficient to get a qualitative understanding
\cite{Wilkins}, but that in a full trimer system the precise monomer eigenenergies may be altered.

In summary, we have shown that by observing the scattering of an incident SP, coupled to sites 1 and 6 of the FMO complex, both the inhibition of the excitonic coupling (excitation-transferring pathway) and entirely missing sites can be detected. We also show that by making the discrete (sites) and the continuous (SPs) channels more disparate, it is possible to further enhance the changes in the transmission spectra of the incident SP. These results provide an alternative method to detect the mutation-induced local defects or the non-functional sites in the FMO complex. We expect that future work will reveal even more uses for such hybrid bio-sensors, including probing the role of quantum coherence and environmental effects in photosynthetic complexes.

\begin{addendum}

\item [Acknowledgement]
This work is supported partially by the National Center for Theoretical
Sciences and Ministry of Science and Technology, Taiwan, grant number  MOST 103-2112-M-006-017-MY4 and MOST 105-2112-M-005-008-MY3. NL and FN acknowledge the support of a grant from the John Templeton Foundation. FN is partially supported by the RIKEN iTHES Project, the MURI Center for Dynamic Magneto-Optics
via the AFOSR award number FA9550-14-1-0040, the IMPACT program of JST, and a Grant-in-Aid for Scientific Research (A).

\item [Author Contributions] YNC and GYC conceived the idea. YAS and MHL carried out the calculations under the guidance of GYC.
All authors contributed to the interpretation of the work and the writing of the manuscript.
\item [Competing Interests]
The authors declare that they have no competing financial interests.

\item [Correspondence]
Correspondence and requests for materials should be addressed to GYC or YNC.
\end{addendum}

\clearpage

%\bigskip

\textbf{Figure 1: Schematic diagram of the system.} Schematic diagram of a monomer of the FMO
complex coupled to nanowire surface plasmons.  A monomer consists of eight (only seven of them are shown here) chromophores. The incident surface plasmon can either be scattered or absorbed, due to the couplings $g_1$ and $g_6$ to the sites $1$ and $6$, respectively. The absorbed excitation then transfers from one chromophore to another, when it arrives at site 3, it can irreversibly transport to the reaction center. \bigskip

\textbf{Figure 2: Transmission spectra of the incident surface plasmon for inhibited pathways.} The red-dashed curves represent the tansmission spectra of the incident surface plasmon when certain excitonic couplings are inhibited: (a) $J_{1,2}$ (b) $J_{2,3}$ (c) $J_{4,7}$ (d) $J_{5,6}$. The black-solid curves represent the transmission spectra of the incident surface plasmon under normal conditions (all the normal excitonic couplings remain). For this figure, we set the dephasing rate $\protect\gamma _{\text{dp}}=77$ cm$^{-1}$, the rate from site 3 to the reaction center $\gamma_s =5.3$ cm$^{-1}$, the SP-site couplings are set to be $g_1=g_6=10$ cm$^{-1}$, and the Ohmic loss rate $\gamma_{\textrm{ol}}$ is set to be 1/20 $\gamma_{\textrm{sp}}$.  \bigskip

\textbf{Figure 3: Transmission spectra of the incident surface plasmon for site-missing.} The red-dashed curves represent the transmission spectra of the incident surface plasmon when sites (a) 2 (b) 5  are disconnected entirely from the FMO complex. The black-solid curves represent the transmission spectra of the incident surface plasmon under normal conditions (all the excitonic couplings remain). For this figure, we set the dephasing rate $\protect\gamma _{\text{dp}}=77$ cm$^{-1}$, the rate from the site 3 to the reaction center $\gamma_s=5.3$ cm$^{-1}$, the SP-site couplings are set to be $g_1=g_6=10$ cm$^{-1}$, and the Ohmic loss rate $\gamma_{\textrm{ol}}$ is set to be 1/20 $\gamma_{\textrm{sp}}$. \bigskip

\textbf{Figure 4: Enhancing changes by asymmetric couplings in transmission spectra for site 2.} The red-dashed curves represent the transmission spectra of the incident surface plasmon for (a) inhibiting the excitonic couplings $J_{1,2}$ (b) disconnecting site 2 from the FMO complex. The black-solid curves represent the transmission spectra of the incident surface plasmon under normal conditions (all the excitonic couplings remain). For this figure, we set the dephasing rate $\protect\gamma _{\text{dp}}=77$ cm$^{-1}$, the rate from the site 3 to the reaction center $\gamma_s=5.3$ cm$^{-1}$, while the SP-site couplings are set to be $g_1=10$ cm$^{-1}$, $g_6=0.1$ cm$^{-1}$, such that $g_1/g_6=100$, and the Ohmic loss rate $\gamma_{\textrm{ol}}$ is set to be 1/20 $\gamma_{\textrm{sp}}$. \bigskip

\textbf{Figure 5: Enhancing changes by asymmetric couplings in transmission spectra for site 5.} The red-dashed curves represent the transmission spectra of the incident surface plasmon when (a) excitonic couplings  $J_{5,6}$ is inhibited  or (b) when site 5 is disconnected entirely from the FMO complex. The black-solid curves represent the transmission spectra of the incident surface plasmon under normal conditions (all the excitonic couplings remain). For this figure, we set the dephasing rate $\protect\gamma _{\text{dp}}=77$ cm$^{-1}$, the rate from the site 3 to the reaction center $\gamma_s =5.3$ cm$^{-1}$, while the SP-site couplings are set to be $g_1=0.1$ cm$^{-1}$, $g_6=10$ cm$^{-1}$, such that $g_1/g_6=0.01$, and the Ohmic loss rate $\gamma_{\textrm{ol}}$ is set to be 1/20 $\gamma_{\textrm{sp}}$. \bigskip

\clearpage
\begin{figure}
\begin{center}
\epsfig{file=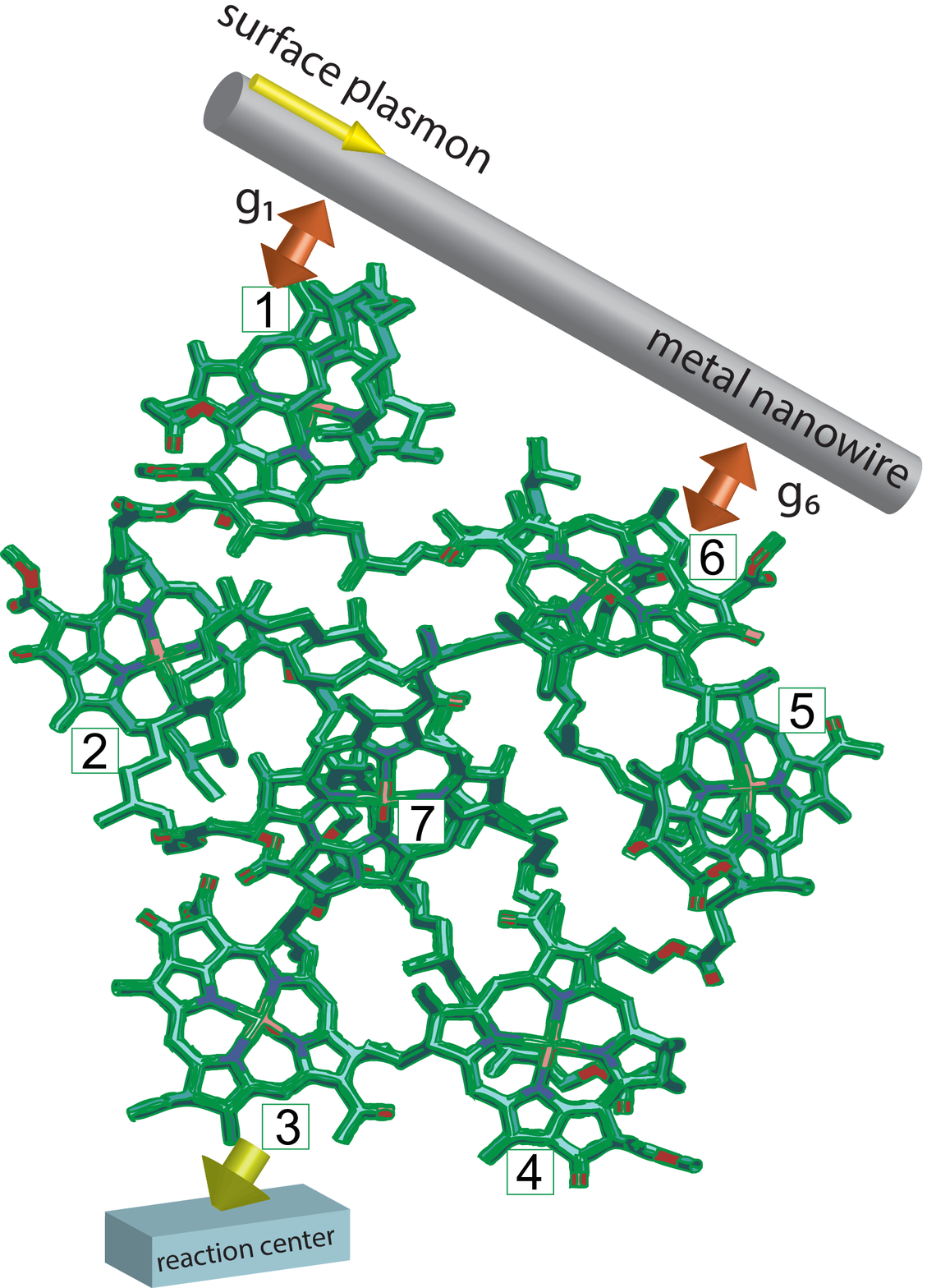,width=7cm}
\end{center}\caption{}
\label{fig1-scheme}
\end{figure}

\clearpage
\begin{figure}
\begin{center}
\epsfig{file=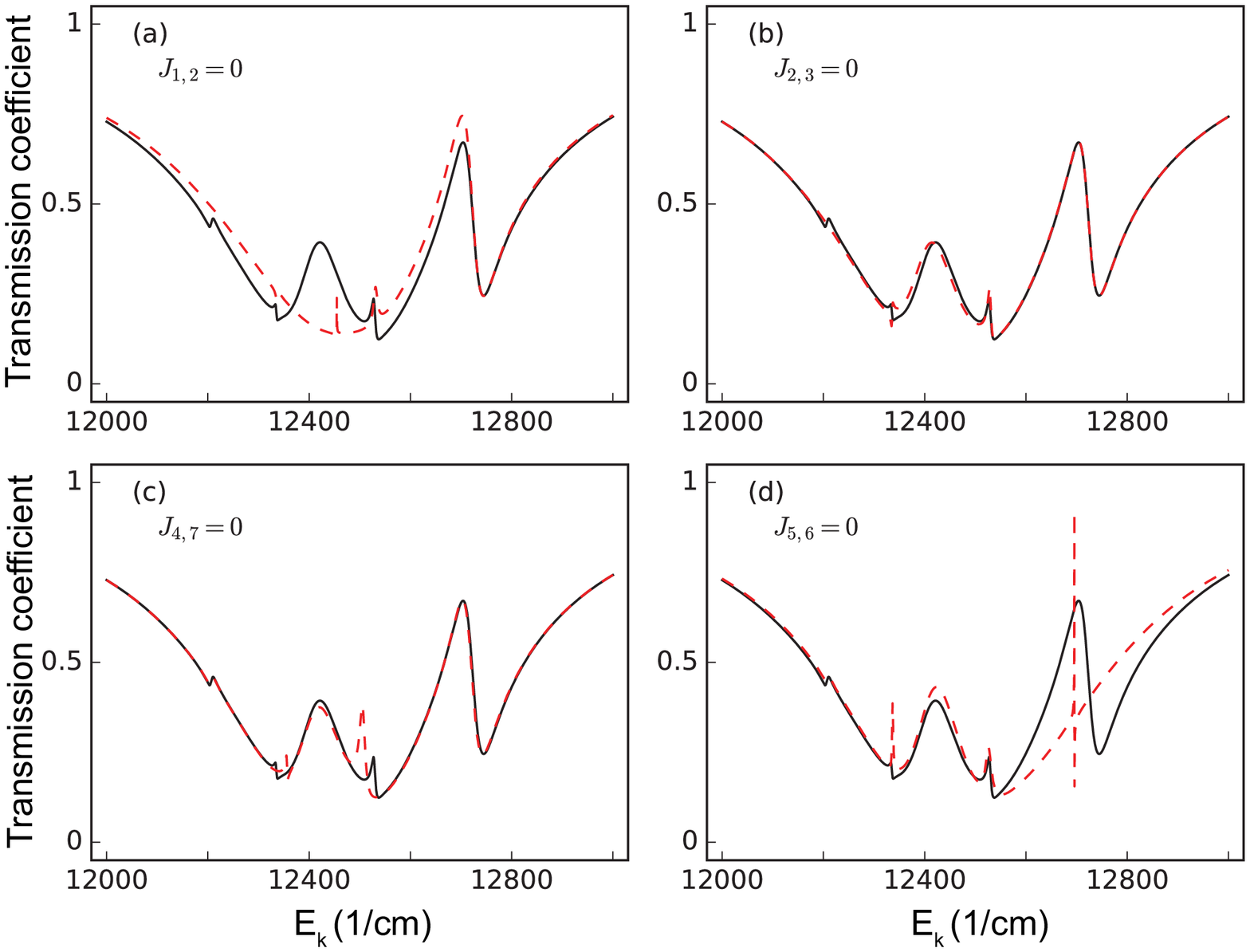,width=10cm}
\end{center}\caption{}
\label{fig2-scheme}
\end{figure}

\clearpage
\begin{figure}
\begin{center}
\epsfig{file=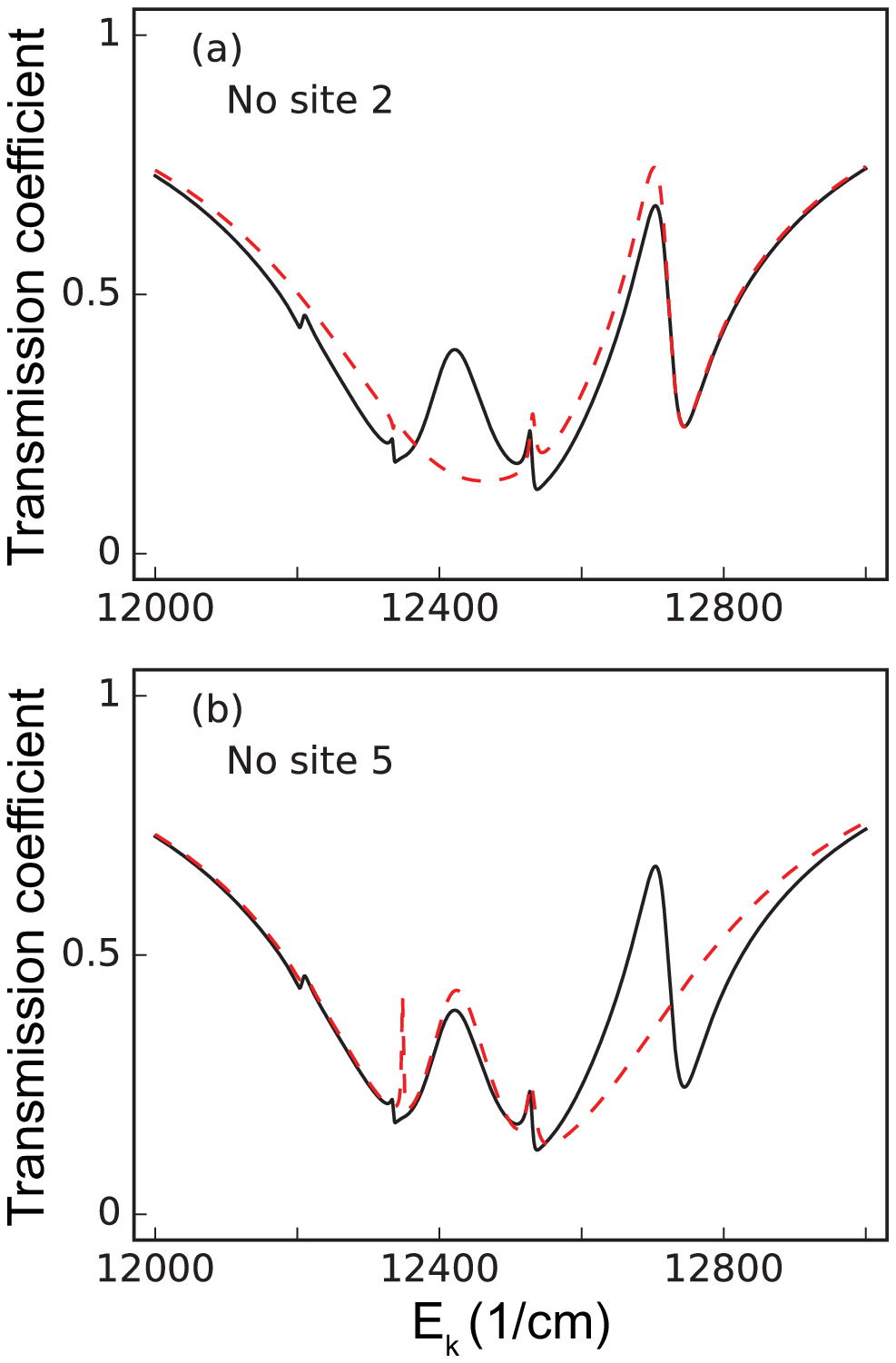,width=10cm}
\end{center}\caption{}
\label{fig3-scheme}
\end{figure}

\clearpage
\begin{figure}
\begin{center}
\epsfig{file=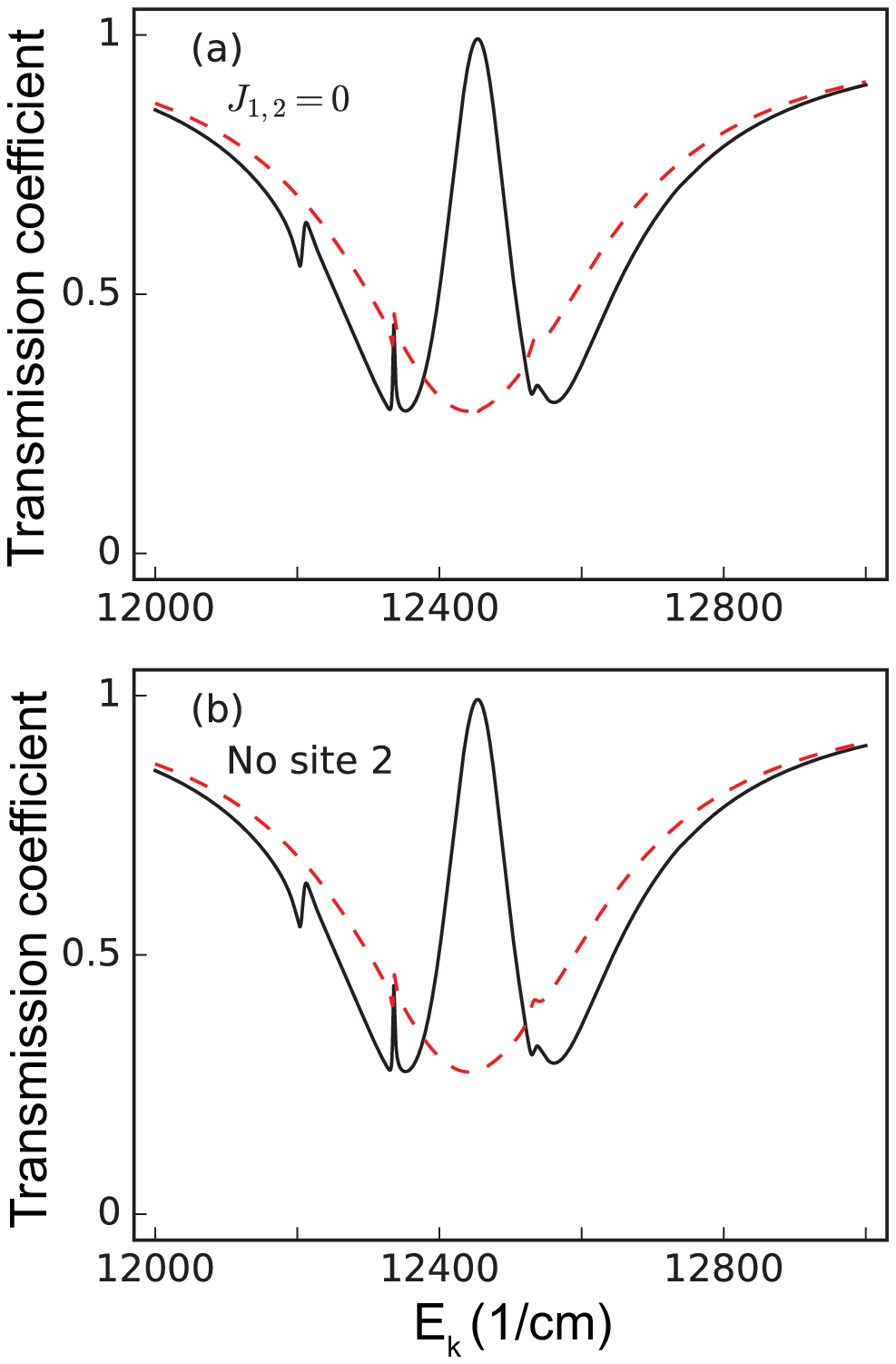,width=10cm}
\end{center}\caption{}
\label{fig4-scheme}
\end{figure}

\clearpage
\begin{figure}
\begin{center}
\epsfig{file=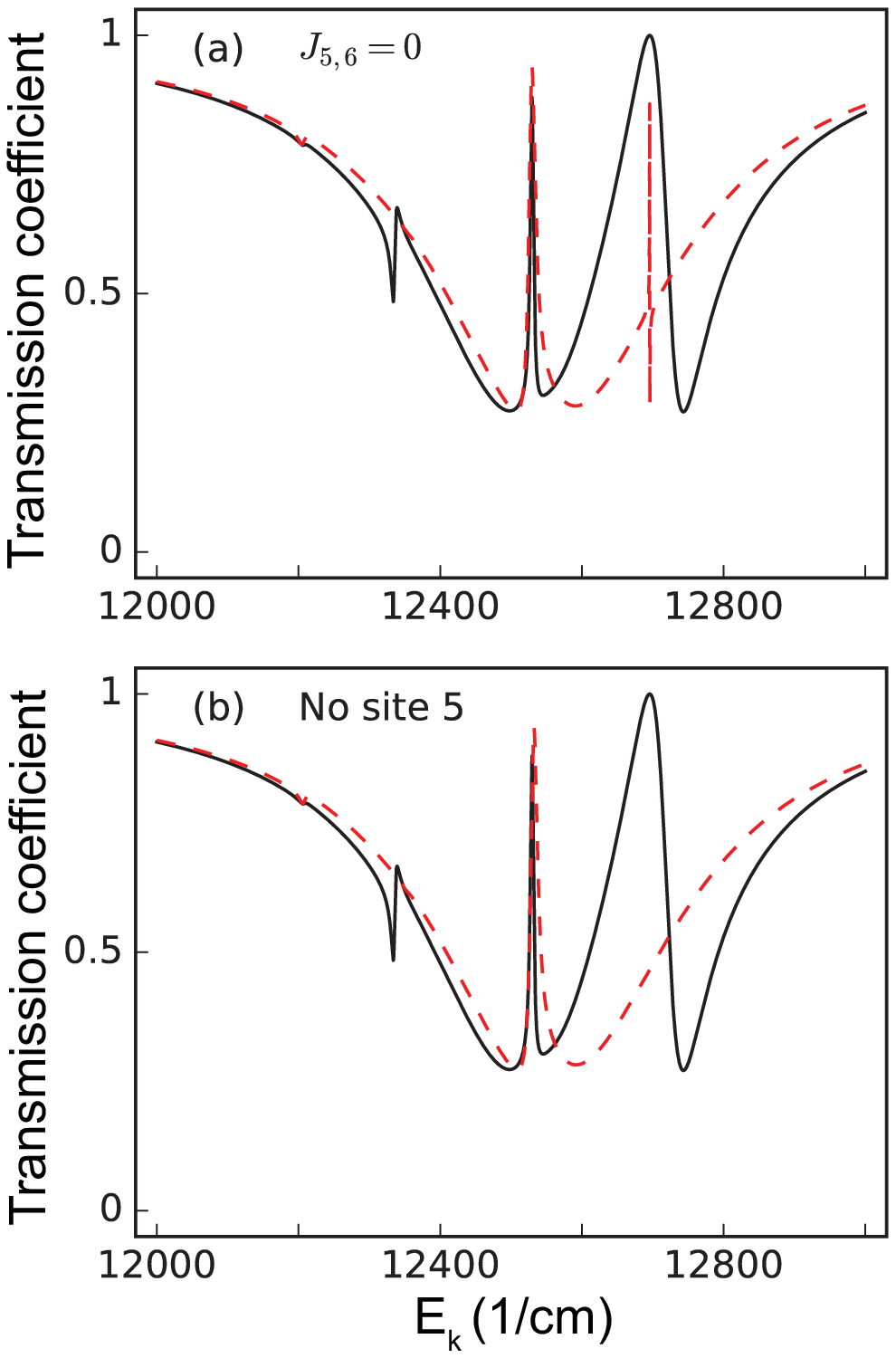,width=10cm}
\end{center}\caption{}
\label{fig5-scheme}
\end{figure}

\end{document}